\documentclass[prc,showpacs,floatfix]{revtex4}
\usepackage{bm}
\usepackage{dcolumn}
\usepackage{graphicx}
\usepackage{amsmath}
\newcommand{\beq}{\begin{equation}}
\newcommand{\eeq}{\end{equation}}
\newcommand{\beqa}{\begin{eqnarray}}
\newcommand{\eeqa}{\end{eqnarray}}
\begin{document}
\title{
A Truncated partial wave analysis of a complete experiment for photoproduction of two
pseudoscalar mesons on a nucleon}
\author{
A.~Fix$^{1}$ and H.~Arenh\"ovel$^{2}$}
\affiliation{
$^1$Laboratory of Mathematical Physics, Tomsk Polytechnic University, Tomsk, Russia\\
$^2$Institut f\"ur Kernphysik, Johannes Gutenberg-Universit\"at Mainz,
D-55099 Mainz, Germany}
\date{\today}
\begin{abstract}
A truncated partial wave analysis for the photoproduction of two
pseudoscalar mesons on a nucleon is discussed with respect to the
determination of a complete set of observables.
For the selection of such a set we have applied a criterion previously
developed for photo- and
electrodisintegration of a deuteron, which allows one to find a 'minimal' set of
observables for determining the partial wave amplitudes up to possible
discrete ambiguities. The question of resolving the remaining ambiguities by invoking additional
observables is discussed for the simplest case, when the partial wave expansion is
truncated at the lowest total angular momentum of the
final state $J_{max}=1/2$.
The resulting 'fully' complete set,
allowing an unambiguous determination of the partial wave amplitudes, is presented.
\end{abstract}

\pacs{13.60.Le, 13.75.-n, 21.45.+v, 25.20.Lj} \maketitle

\section{Introduction}

Photoproduction of two pseudoscalars on nucleons have been studied rather
intensively during the last two decades. At present a large amount of
experimental data, in particular for $\pi^0\pi^0$ and $\pi^0\eta$
channels, have been collected and some new experiments on polarization
observables for these reactions are planned.
The interpretation of the data within different models has allowed one to qualitatively
understand the major mechanisms of these processes. At the same time, although the
general agreement of the various calculations with the measured cross
sections is reasonable,
significant qualitative differences between these models exist. Some
of them were already discussed, for example in Ref.~\cite{Kashev2pi0}.

One reason for these differences between theoretical results is that the standard
approach, based on the isobar model, appears to have reached certain limitations.
Probably its weakest point is that the corresponding formalism depends on a specific
assumption about the dynamics of the production process. Within the
isobar model approach
one assumes that the two-body discontinuity in the reaction matrix coming from the
interaction in the two-body subsystems in the final state may be approximated by
resonance terms (usually taken in the Breit-Wigner form). In other words, it is assumed
that the final three-particle state is produced via intermediate formation of quasi
two-body states containing meson-nucleon and meson-meson isobars.

In order to achieve a significant improvement of present theoretical models one
has to eliminate as much as possible the mentioned model dependencies from the formal
description of these reactions. In this respect, an ideal tool for the investigation of
the reaction dynamics is an analysis of a complete experiment within a
given model,
based on the fundamental principles of rotation and parity invariance. It is clear that
for the photoproduction of two mesons this task is considerably more
complicated in comparison to the
photoproduction of a single meson. Firstly, in the case of
three particles in the final state, the amplitudes depend on five kinematical variables, so that for
their determination accurate measurements of five-dimensional distributions are needed. Secondly, 
contrary to the single meson case, a complete set contains  a
considerably larger number of observables. Indeed, if two mesons are produced, the property of parity
conservation does not allow one to reduce the number of
independent amplitudes. Therefore, in order to determine all eight complex
amplitudes up to an overall
phase one needs at least 15 observables. As was shown by Roberts and
Oed \cite{Roberts}, a complete set includes not only single and double
but also triple polarization observables. The latter is especially
disappointing since it requires the implementation of complicated
measurements at a high level of accuracy.

At the same time, as was noted in Refs.~\cite{Workman} and
\cite{Tiat}, arbitrariness in the overall phase at each
point of the phase space does not allow one to find the 
multipole amplitudes, which are obviously needed for a nucleon
resonance analysis.
Therefore, if one searches for resonances or, more generally, for
states with definite spin and parity $J^P$, it is more reasonable (and probably less
complicated) to adopt a truncated partial wave expansion up to a maximal total angular
momentum $J_{max}$ and to study instead of the spin or helicity amplitudes the partial
wave amplitudes. In this case, as a rule, a lower number of polarization observables is
needed. As is discussed in Refs.~\cite{Tiat,Grush,Omel} for $\gamma
N\to \pi^0 N$, in order
to determine (up to an overall phase) the multipoles $E_{0+}$, $E_{1+}$, and $M_{1\pm}$
which are important in the first resonance region, already a set of single polarization
observables with an additional measurement of only one of the double
polarization observables (for example, $F$- or $G$-asymmetry) is sufficient.

For a partial wave analysis we firstly need a convenient and model
independent form of the
partial wave decomposition of the reaction amplitude. As already noted above,
the isobar model does not meet this requirement,
since it depends on specific assumptions about the reaction dynamics. Therefore,
we adopt in the present paper the formalism developed in
Ref.~\cite{FiA12}. Here a special
coordinate frame is used in which the $z$-axis is chosen along the normal to the plane
spanned by the momenta of the final particles in the overall center-of-mass frame. In this case one can
choose as independent
kinematical variables the energies of two of the three final particles and three
angles, determining the orientation of the final state plane with respect to the photon
beam (one of the angles corresponds to the rotation around the normal). Then the
amplitude for a given total angular momentum $J$ of the final state and its
projection $M$ on the $z$-axis is obtained by an expansion of the helicity
amplitudes over a set of Wigner functions. In Ref.~\cite{Kashev2pi0} this approach was
applied successfully to the analysis of the unpolarized differential cross section for
the photoproduction of $\pi^0\pi^0$ pairs.

The second important question is what is an optimal choice of the
observables needed for the determination of the partial wave
amplitudes. While for the determination of $n$ complex
quantities (up to an overall phase factor) only $2n-1$ real parameters are needed, the
total number of linearly independent observables (or, more generally, the number of
bilinear hermitian forms of the amplitudes) is $n^2$. This means that there exist also
nonlinear dependencies between the observables, so that not any subset of
$2n-1$ observables may form a complete set from which the $2n-1$ real
parameters can be deduced.

Thus the question is how to select out of this set of $n^2$
observables  a subset of $2n-1$ independent observables.
In the present paper we describe a method
which may be used to single out a minimal set of independent
observables. The criterion, which underlies the method,
was originally developed for deuteron photo- and electrodisintegration in
Refs.~\cite{ArLeiTom,ArLeiTomFBS}. An additional very important
question concerns possible discrete ambiguities, which naturally can
appear since the extraction of the amplitudes implies a solution of a
set of quadratic equations. To resolve these ambiguities we adopt for
our truncated partial wave analysis an additional criterion, which was
used in Ref.~\cite{Tabakin} for the spin amplitude analysis of single
meson photoproduction.

First, we present in the next section a brief review of the
formalism developed in Ref.~\cite{FiA12}. Then we give in
Sec.~\ref{truncatedpwa} an example of the method of how
to find a complete set of observables for the simplest case, when only partial
waves with $J=1/2$ and both parities are included. In
  Sect.~\ref{conclusion} we summarize our results and give an outlook
on future developments. Some details are collected in two appendices.

\section{Formal developments}\label{formal}

\begin{figure}[h]
\begin{center}
\includegraphics[scale=.8]{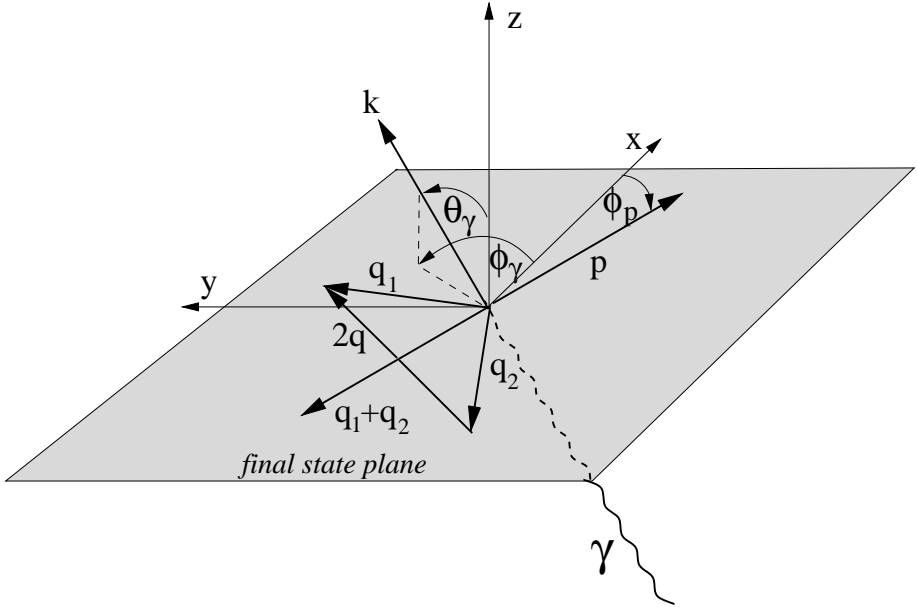}
\caption{Definition of the coordinate system in the c.m.\ system. } \label{fig1}
\end{center}
\end{figure}

For the theoretical description we choose the overall c.m.~system. The four-momenta of
incoming photon, outgoing mesons, and final nucleon are denoted by
$(\omega_\gamma,\vec{k}\,)$, $(\omega_1,\vec{q}_1\,)$, $(\omega_2,\vec{q}_2\,)$,
and $(E,\vec{p}\,)$, respectively. We consider within this system
two right-handed orthogonal coordinate systems: (i) one associated with the incoming
photon called $K_\gamma$ with $z$-axis along the photon momentum and
$x$-axis arbitrary, and
(ii) the so-called ``rigid body'' system $K_{fs}$, associated with the final state plane
spanned by the final three particles, in which the $z$-axis is taken to be the normal to
this plane and parallel to $\vec p\times\vec q_1$. Thus the  $x$- and $y$-axes are in the
final scattering plane (see Fig.~\ref{fig1}). The transformation from $K_{fs}$  to
$K_\gamma$ is given by a rotation through Euler angles $(\phi,\theta,0)$. Thus
relative to $K_{fs}$ the photon momentum $\vec{k}$ has the spherical angles
$\Omega_\gamma=(\theta_\gamma,\phi_\gamma)$.

For the $T$-matrix we had derived in Ref.~\cite{FiA12} the following
expression expanding the final state into partial waves
\beqa\label{14}
T_{\nu \lambda\mu}(\phi_p,\omega_1,\omega_2, \Omega_\gamma)&=&
e^{-i\nu\phi_p}\sum\limits_{JM_J} t^{JM_J}_{\nu\lambda\mu}(\omega_1,\omega_2)
\,D^J_{M_J\,\lambda-\mu}(R_{\gamma p}) \,,
\eeqa %
where $J$ and $M_J$ denote respectively the total angular momentum of
the partial wave and its projection on the normal to the final state plane.
The rotation matrix $D^J_{M'M}$ is taken in the convention of
Rose~\cite{Ros57} with argument $R_{\gamma p}=(\phi_{\gamma
  p},\theta_\gamma,-\phi_{\gamma})$ and $\phi_{\gamma
  p}=\phi_{\gamma}-\phi_{ p}$.  The helicities of photon and initial
and final nucleons are denoted by $\lambda$, $\mu$, and $\nu$,
respectively. As independent variables we had chosen besides
the photon angles $\Omega_\gamma=(\theta_\gamma,\phi_\gamma)$  and the proton angle
$\phi_p$ in the final state plane, the energies of the two final mesons $\omega_1$ and
$\omega_2$. The latter two determine the relative angle
$\phi_{qp}=\phi_q-\phi_p$ between the momentum of the final proton and
the relative momentum $\vec q = (\vec q_1-\vec q_2)/2$ of the two
mesons according to
\beq\label{cosphi_qp}
\cos\phi_{qp}=
\frac{1}{2qp}(\omega_2^2-\omega_1^2-M_2^2+M_1^2)\,,
\eeq
with the final nucleon momentum
\beq\label{nucleon_p}
p=|\vec{p}\,|=\sqrt{(W-\omega_1-\omega_2)^2-M_N^2}\,,
\end{equation}
where $W$ denotes the invariant total energy,
and  the relative momentum $q$ of the two mesons is determined by
\beq \label{meson_q}
q^2=\frac{1}{2}(\omega_1^2+\omega_2^2-M_1^2-M_2^2)-\frac{p^2}{4}\,.
\eeq
In the foregoing equations the masses of the nucleon and the two mesons
are denoted by $M_N$, $M_1$ and $M_2$, respectively.

The expression in Eq.~(\ref{14}) is obtained by
making use of rotation and inversion invariance (that is angular momentum and parity
conservation) and is therefore completely general. The final partial
wave is taken in the form $ |q p;\, ((l_p \frac{1}{2})j_p l_q)J M_J
\rangle ^{(-)}$ where $l_p$ and $l_q$ denote the angular momenta of
nucleon and meson pair, respectively.

The contribution of the final partial wave to the reaction amplitude is given by
\begin{eqnarray}
t^{JM_J}_{\nu \lambda\mu}(\omega_1,\omega_2)&=&
t^{JM_J}_{\nu \lambda\mu} (\phi_{qp})\nonumber\\
&=&\sum_{l_p j_p m_p L}
\left(\begin{array}{ccc} l_p &
    \frac{1}{2} & j_p \cr 0 & \nu & -\nu\cr
\end{array}\right)
\left(\begin{array}{ccc} J & L & \frac{1}{2} \cr \mu-\lambda & \lambda & -\mu\cr
\end{array}\right)
d^{j_p}_{\nu\, m_p}(\pi/2) \,
e^{i(M_J-m_p)\phi_{qp}}\,{\cal O}^{\lambda LJ}_{M_J}(l_p j_p m_p)\,,\label{multipole}
\end{eqnarray}
with
\beqa
{\cal O}^{\lambda LJ}_{M_J}(l_p j_p m_p) &=&
\frac{(-1)^{1+J}\widehat J}{2\sqrt{2\pi}} \sum_{l_q m_q} i^L (-1)^{l_p+j_p+l_q}
\,\widehat l_p\,\widehat j_p\,\widehat l_q\,\widehat L \,d^{l_q}_{0\,
m_q}(\pi/2)\nonumber\\&&\times
\left(
\begin{matrix}
j_p& l_q&J \cr m_p&m_q&-M_J \cr
\end{matrix} \right)
\langle p \,q; \big((l_p\frac12)j_p l_q\big)J||{\cal O}^{\lambda
  L}||\frac{1}{2}\rangle\,,
\eeqa
where $\hat l=\sqrt{2l+1}$ and ${\cal O}^{\lambda  L}$ denotes the
electromagnetic multipole operator with electric and magnetic
contributions
\begin{eqnarray}
{\cal O}^{\lambda L}_{M}&=& E_{M}^L +\lambda M_{M}^L\,.
\end{eqnarray}

The amplitudes $t_{\nu\lambda\mu}^{JM_J}$ obey the following symmetry
property  which follows from parity conservation (see
Ref.~\cite{FiA12}, one should note a misprint in the phase)
\begin{equation}
t_{-\nu -\lambda-\mu}^{JM_J}(\phi_{qp})
=(-1)^{\nu-M_J}t_{\nu\lambda\mu}^{J-M_J}(-\phi_{qp})\,.\label{sym-t}
\end{equation}
The interchange $\phi_{qp}\leftrightarrow -\phi_{qp}$ corresponds to
the interchange $(\omega_1,M_1)\leftrightarrow (\omega_2,M_2)$, see
Eq.~(\ref{cosphi_qp}) through
(\ref{meson_q}). Therefore, parity conservation does not reduce the
number of independent amplitudes in this case in contrast to single
meson production as already noted in Ref.~\cite{Roberts}. However, in
a more general sense this relation allows one to reduce the number of
independent amplitudes. Namely, provided that the amplitude
$t_{\nu\lambda\mu}^{JM_J}$ is known in the whole region of the Dalitz
plot $(\omega_1,\omega_2)$ the amplitude $t_{-\nu-\lambda-\mu}^{J-M_J}$
can be obtained using the symmetry relation of Eq.~(\ref{sym-t}). Thus in
this more general sense one has $4(2J+1)$ independent amplitudes for
a given $J$, except for $J=1/2$ for which this number
is reduced to $2(2J+1)=4$ because of an additional requirement
$\lambda-\mu\leq J$, coming from angular momentum conservation.

As shown in Appendix A, one can separate the contributions of those final states with
positive parity from those of negative one according to
\beq\label{parity1}
t^{JM_J}_{\nu
\lambda\mu}=(-)^{\delta_{\nu -\frac{1}{2}}(\frac{1}{2}+M)} (t^{JM_J\,+}_{\lambda\mu}+2\nu\,
t^{JM_J\,-}_{\lambda\mu})\,, \eeq where
\beq\label{parity1a}
t^{JM_J\,\pm}_{\lambda\mu}=\frac{1}{2}\Big(t^{JM_J}_{1/2
  \lambda\mu}\pm(-)^{\frac{1}{2}+M_J} t^{JM_J}_{-1/2 \lambda\mu} \Big)\,.
\eeq

An interesting consequence of Eq.~(\ref{parity1}) is that the bilinear expression
$t^{J'M'_J*}_{\nu'\lambda'\mu'}t^{JM_J}_{\nu\lambda\mu}$ with
$\nu'=\nu$ is invariant
under the transformation (parity exchange)
\beq
t_{\lambda\mu}^{JM_J+}\leftrightarrow
t_{\lambda\mu}^{JM_J-}\,,\label{paritytrans}
\eeq
 which, as follows from Eq.\,(\ref{parity1a}),
is equivalent to the transformation of the matrices $t^{JM_J}_{\nu\lambda\mu}$
\beq\label{paritytrans1}
t^{JM_J}_{\nu\lambda\mu}\rightarrow 2\nu\, t^{JM_J}_{\nu\lambda\mu}\,.
\eeq
Indeed, one finds successively (note $2\nu=\pm 1$)
\begin{eqnarray}
t^{J'M'_J*}_{\nu\lambda'\mu'}t^{JM_J}_{\nu\lambda\mu}&=&(t^{J'M_J'\,+}_{\lambda'\mu'}+2\nu\,
t^{J'M'_J\,-}_{\lambda'\mu'})^*(t^{JM_J\,+}_{\lambda\mu}+2\nu\, t^{JM_J\,-}_{\lambda\mu}) \nonumber\\
&=& (2\nu\,t^{J'M_J'\,+}_{\lambda'\mu'}+
t^{J'M'_J\,-}_{\lambda'\mu'})^*(2\nu\,t^{JM_J\,+}_{\lambda\mu}+t^{JM_J\,-}_{\lambda\mu}) \nonumber\\
&\Rightarrow& (2\nu\, t^{J'M'_J\,-}_{\lambda'\mu'}+t^{J'M_J'\,+}_{\lambda'\mu'})^*(2\nu\,
t^{JM_J\,-}_{\lambda\mu}+t^{JM_J\,+}_{\lambda\mu})
=t^{J'M'_J*}_{\nu\lambda'\mu'}t^{JM_J}_{\nu\lambda\mu}\,.
\end{eqnarray}
Therefore, in order to distinguish between the
contributions of states with different parities one has to measure recoil
polarization along the $x$ or $y$ axes which are governed by such
bilinear expressions with $\nu'\neq\nu$.

As observables we will consider the differential cross section and the recoil
polarization for unpolarized and circularly polarized photons. As shown
in Ref.~\cite{ArF13} the differential cross section with circular beam
asymmetry is given by
\beq\label{difcrossec}
\frac{d^5\sigma(P^\gamma_c)}{
d\phi_pd\omega_1d\omega_2d\Omega_\gamma}= \frac{d\sigma_0}{
d\phi_pd\omega_1d\omega_2d\Omega_\gamma} \Big(1+P^\gamma_c\,\Sigma^c
\Big)\,,
\eeq
where $P^\gamma_c$ denotes
the degree of circular polarization. The unpolarized differential cross section is
\beq
\frac{d\sigma_0}{ d\phi_p
d\omega_1d\omega_2d\Omega_\gamma}=T^{0}_{00}\,, \label{dsig0}
\eeq
and the beam asymmetry for circular photon polarization
\beq
\Sigma^c\,T^{0}_{00}=T^{c}_{00}\,,\label{dsigasyc}
\eeq
where
\beq
T^{0/c}_{00}=
c(W)\,\mathrm{Re} \,v_{00;00}^{0/c}
\,\label{tau0c}
\end{equation}
with $c(W)=M_N^2/(4(2\pi)^4(W^2-M_N^2))$ as a kinematical factor.
Here the quantities $v^{ \alpha}_{I'M';IM} $ with
$\alpha\in\{0,c\}$ are defined by
\beqa
v^{ 0}_{I'M';IM} &=& \frac{1}{1+\delta_{M0}}\,
\sum_{\lambda} u_{I'M';IM}^{\lambda\lambda}\,,\label{v0}\\
 v^{c}_{I'M';IM} &=& \frac{1}{1+\delta_{M0}}\,
\sum_{\lambda}\lambda\, u_{I'M';IM}^{\lambda\lambda}\,,\label{vc}
\eeqa
where the $u_{I'M';IM}^{\lambda'\lambda}$ contain bilinear combinations of the
partial wave amplitudes $t^{JM_J}_{\nu
  \lambda\mu}$, given in Eq.~(\ref{multipole}), according to
\beqa\label{uLL}%
u_{I'M';IM}^{\lambda'\lambda}&=& (-)^{M}\,\widehat {I'}\widehat I\,
\sum_{ j m'm}(2j+1) \,D^j_{m'm}(R_{\gamma p}) \nonumber\\
&&\hspace{.5cm}\times
\sum_{\nu'\nu \mu' \mu} (-1)^{\nu'}
\left(
\begin{matrix}
\frac{1}{2}& \frac{1}{2}&I' \cr \nu&-\nu'&M' \cr
\end{matrix} \right)
\left(
\begin{matrix}
\frac{1}{2}& \frac{1}{2}&I \cr \mu&-\mu'&-M \cr
\end{matrix} \right)
\nonumber\\
&&\hspace{.5cm}\times\sum_{J'M_J' JM_J} (-1)^{-M_J}
\left(
\begin{matrix}
J'& J&j \cr M_J'&-M_J&m' \cr
\end{matrix} \right)
\left(
\begin{matrix}
J'& J&j \cr \lambda'-\mu'&\mu-\lambda&m \cr
\end{matrix} \right)
\nonumber\\ &&\hspace{2cm}
\times\, t^{J'M_J'}_{\nu'
  \lambda'\mu'}(\omega_1,\omega_2)^*
\,t^{JM_J}_{\nu \lambda\mu}(\omega_1,\omega_2)
\,,
\eeqa%
with $R_{\gamma p}=(\phi_{\gamma p},\theta_\gamma,- \phi_{\gamma })$. These quantities
have the following symmetry property:
\beq u_{I'M';IM}^{\lambda\lambda'}=(-1)^{M'+M}(u_{I'-M';I-M}^{\lambda'\lambda})^*\,.
\label{symma}
\eeq%
One should note that in Eqs.~(\ref{v0}) and (\ref{vc}) only
$u_{I'M';IM}^{\lambda'\lambda}$ with $\lambda'=\lambda$ appear.
For the recoil polarization component $P_{x_i}$ of the outgoing
nucleon one has 
\beq\label{recoilpol}
P_{x_i}\frac{d^5\sigma(P^\gamma_c)}{
d\phi_pd\omega_1d\omega_2d\Omega_\gamma}= \frac{d\sigma_0}{
d\phi_pd\omega_1d\omega_2d\Omega_\gamma} \Big( P^0
_{x_i}+P^\gamma_c\,P^c_{x_i}\Big)
\eeq
with recoil polarizations for unpolarized beam and target
\beq
P^0 _{x_i} \,T^{0}_{00}=P^{x_i,0}_{00}\,, \label{recoilp0}
\eeq
as well as beam asymmetries for circularly polarized photons
\beq
P^c_{x_i}\,T^{0}_{00}=P^{x_i,c}_{00}\,,\label{recoilpc}\\
\eeq
where for $\alpha\in\{0,c\}$
\begin{eqnarray}
 (P/Q)^{ x,\alpha}_{00}&=&-\sqrt{2}\,c(W)\,
\mathrm{Re/Im}\, w_{11;00}^{\alpha,+}
\,,\label{asyx}
\\
 (P/Q)^{ y,\alpha}_{00}&=&\mp\sqrt{2}\, c(W)\,
\mathrm{Im/Re}\, w_{11;00}^{\alpha,-}
\,,\label{asyxy}
\\
 (P/Q)^{ z,\alpha}_ {00}&=&c(W)\,
\mathrm{Re/Im}\, w_{10;00}^{\alpha,+} \,,
\label{asyz}
\end{eqnarray}
with
\beq
w^{\alpha,\pm}_{I'M';IM} = \frac{1}{2}\,
(e^{iM'\phi_p}\,v^{\alpha}_{I'M';IM}
  \pm (-)^{M'}\,e^{-iM'\phi_p}\,\,v^{\alpha}_{I'-M';IM})\,.\label{walpha}
\eeq
This completes the formal part. 

\section{A Truncated partial wave analysis}\label{truncatedpwa}

In this section we consider a method which allows one to find  for a
reaction with $n$ independent complex amplitudes a complete subset of
$2n-1$ independent observables. It was
developed in Refs.~\cite{ArLeiTom,ArLeiTomFBS} and applied to the
analysis of deuteron electro- and
photodisintegration, and we refer the reader to this paper for more
details. At first, in order to explain the key points of the method, we
consider a very simple mathematical example. Given a
2-dimensional real vector $\vec{x}=\{x_1,x_2\}$, whose components are
called ``amplitudes'', and two real quadratic forms $f_k$, called ``observables'',
\begin{equation}\label{QuadForms}
f_k(\vec{x}\,)=\frac{1}{2}\sum_{ij=1}^2x_iA^k_{ij}x_j\,,\quad
k=1,2\,,
\end{equation}
where the two matrices $A^k$ are symmetric,
then the question is under which conditions for the matrices one can
determine the amplitudes $\{x_1,x_2\}$ from given values $\{f_1,f_2\}$
of the observables.
In other words, what is the criterion, that the set of quadratic equations
(\ref{QuadForms}) can be inverted (apart from possible quadratic ambiguities).

For the moment being, let us assume that
$\vec{x}^{\,0}=\{x^0_1,x^0_2\}$ is the required solution of
Eq.~(\ref{QuadForms}) for given values $\{f_1,f_2\}$. A necessary
condition for the inversion is that in the
neighborhood of $\vec{x}^{\,0}$ the Jacobian of the transition $\{x_1,x_2\}\to
\{f_1,f_2\}$ is nonvanishing, i.e.\ using Eq.~(\ref{QuadForms})
\beqa\label{cond}
\det\left(\frac{\partial f_k}{\partial x_i}\right) &=&\det\left(
\begin{array}{cc}
A^1_{11}x_1+A^1_{12}x_2 & A^2_{11}x_1+A^2_{12}x_2 \\
 & \\
A^1_{21}x_1+A^1_{22}x_2 & A^2_{21}x_1+A^2_{22}x_2
\end{array}\right)\nonumber\\
&=&x_1^2 \det(\widetilde{A}^1)+x_1x_2 \det(\widetilde{A}^2)+ x_2x_1
\det(\widetilde{A}^3)+x_2^2 \det(\widetilde{A}^4)\neq 0\,.
\eeqa
Here the new matrices $\widetilde{A}^i$ ($i=1,\dots ,4$) are
constructed as all possible combinations of the columns of the initial
matrices ${A}^k$.

The condition in Eq.~(\ref{cond}) now reads: {\it if the Jacobian
  $\det (\frac{\partial
f_k}{\partial x_i})$ is nonvanishing, then at least one of the determinants
$\det\widetilde{A}_i$ ($i=1,\dots, 4$) is nonvanishing}. This statement can be
reformulated as a sufficient condition for the degeneracy of the
transition $\{x_1,x_2\}\to
\{f_1,f_2\}$. Namely, {\it if all determinants $\det\widetilde{A}^i$
  ($i=1,\dots,4$) vanish, than the set of quadratic equations
  (\ref{QuadForms}) cannot be inverted.}

Now we would like to apply the above criterion to the reaction with
two pseudoscalar mesons in the final state. As already mentioned, we
perform a truncated partial wave analysis, where the amplitude is
decomposed over the partial wave amplitudes up to some
maximum value of the total angular momentum $J_{max}$.
As a set of observables for the truncated partial wave analysis it is
convenient to choose real and
imaginary parts of the coefficients appearing in the expansion of the the functions
$u_{IM,I^\prime M^\prime}^{\lambda'\lambda}$, defined in
Eq.~(\ref{uLL}) over the Wigner functions. Rewriting Eq.~(\ref{uLL}) as
\begin{equation}
u_{I'M';IM}^{\lambda'\lambda}= \sum_{j=0}^{2J_{max}}\sum_{m'm}
U_{jm'm}^{I'M'IM,\lambda'\lambda}(\omega_1,\omega_2)D_{m'm}^j(R_{\gamma p})\,,
\end{equation}
one obtains the observables $U_{jm'm}^{I'M'IM,\lambda'\lambda}$ in
terms of bilinear combinations of the partial wave amplitudes
$t_{\nu\lambda\mu}^{JM_J}$ 
\begin{eqnarray}\label{Walpha}
U_{jm'm}^{I'M'IM,\lambda'\lambda}(\omega_1,\omega_2)&=&(-1)^{M}\,\widehat {I'}\widehat
I\,
(2j+1)\sum_{\nu'\nu \mu' \mu} (-1)^{\nu'} \left(
\begin{matrix}
\frac{1}{2}& \frac{1}{2}&I' \cr \nu&-\nu'&M' \cr
\end{matrix} \right)
\left(
\begin{matrix}
\frac{1}{2}& \frac{1}{2}&I \cr \mu&-\mu'&-M \cr
\end{matrix} \right)
\nonumber\\
&\times&\sum_{J'M_J' JM_J} (-1)^{-M_J} \left(
\begin{matrix}
J'& J&j \cr M_J'&-M_J&m' \cr
\end{matrix} \right)
\left(
\begin{matrix}
J'& J&j \cr \lambda'-\mu'&\mu-\lambda&m \cr
\end{matrix} \right)
\nonumber\\ &\times&\, t^{J'M_J'}_{\nu'
  \lambda'\mu'}(\omega_1,\omega_2)^*
\,t^{JM_J}_{\nu \lambda\mu}(\omega_1,\omega_2)\,.
\end{eqnarray}
The symmetry property of Eq.~(\ref{symma}) leads to the following
symmetry of the observables for the interchange $\lambda\leftrightarrow\lambda'$
\begin{equation}\label{165}
U_{jm'm}^{I'M'IM,\lambda\lambda'}=(-1)^{M'+M+m'+m}(U_{j-m'-m}^{I'-M'I-M,\lambda'\lambda})^*.
\end{equation}
For the discussion to follow it is convenient to introduce a matrix
notation by writing
\begin{equation}
U_{jm'm}^{I'M'IM,\lambda'\lambda}=\sum_{k'k=1}^nt^*_{k'}U^{(\beta,\lambda'\lambda)}_{k'k}t_k\,,
\end{equation}
where $\beta=(I'M';IM;jm'm)$ and
$k^{(\prime)}=(J^{(\prime)},M^{(\prime)}_J,\nu^{(\prime)},\mu^{(\prime)})$
enumerates the amplitudes. The maximum
value $n$ of the indices $k'$ and $k$ is equal to the total number of amplitudes
$t^{JM_J}_{\nu\lambda\mu}$ for a given value of $\lambda$. The symmetry
relation in Eq.~(\ref{165}) leads to the matrix relation
\begin{equation}\label{167}
U^{(I'M';IM;jm'm,\lambda'\lambda)}=(-1)^{M'+M+m'+m}(U^{(I'-M';I-M;j-m'-m,\lambda\lambda')})^T\,.
\end{equation}
Furthermore, using the property
\begin{equation}
\sum_m(-1)^{j-m}
\left(
\begin{matrix}
j& j& J \cr m & -m & 0 \cr
\end{matrix} \right)=\sqrt{2j+1}\,\delta_{J0}\,,
\end{equation}
one obtains for the trace
\begin{eqnarray}
\sum_k
U^{(\beta,\lambda'\lambda)}_{kk}&=&\delta_{M'0}\delta_{M0}\delta_{m'0}
\delta_{m,\lambda-\lambda'}\widehat {I'}\widehat
I\,(2j+1) \sum_{JM_J}
(-1)^{-M_J} \left(
\begin{matrix}
J & J&j \cr M_J&-M_J& 0 \cr
\end{matrix} \right)
\nonumber\\
&&\times
\sum_{\nu} (-1)^{\nu} \left(
\begin{matrix}
\frac{1}{2}& \frac{1}{2}&I' \cr \nu&-\nu & 0 \cr
\end{matrix} \right)
\sum_{\mu} \left(
\begin{matrix}
\frac{1}{2}& \frac{1}{2}&I \cr \mu&-\mu & 0 \cr
\end{matrix} \right)
\left(
\begin{matrix}
J & J&j \cr \lambda'-\mu &\mu-\lambda& \lambda-\lambda'\cr
\end{matrix} \right) \nonumber\\
&=& (2J_{max}+1)\delta_{I'0}\delta_{M'0}\delta_{I0}\delta_{M0}\delta_{j0}\delta_{m'0}\delta_{m0}\delta_{\lambda'\lambda}\,,
\end{eqnarray}
which means that all matrices
$U^{(I'M';IM;jm'm,\lambda'\lambda)}$ have a vanishing trace except for the
diagonal matrix $U^{(00;00;000;\lambda\lambda)}$.

The real and imaginary parts of the coefficients
$U_{jm'm}^{I'M'IM,\lambda'\lambda}=f^{(\beta,\lambda'\lambda)} +
i\,g^{(\beta,\lambda'\lambda)}$ may
now be treated as observables for the truncated partial wave
analysis.
Again we introduce a matrix representation by
\beqa\label{ReU}
f^{(\beta,\lambda'\lambda)}&=&\mathrm{Re}\,U_{jm'm}^{I'M'IM,\lambda'\lambda}=
\frac{1}{2}\sum_{k'k=1}^nt^*_{k'}F^{(\beta,\lambda'\lambda)}_{k'k}t_k\,,\\
g^{(\beta,\lambda'\lambda)}&=&\mathrm{Im}\,U_{jm'm}^{I'M'IM,\lambda'\lambda}
=\frac{1}{2i}\sum_{k'k=1}^nt^*_{k'}G^{(\beta,\lambda'\lambda)}_{k'k}t_k\,,
\label{ImU}
\eeqa
where the matrices $F^{(\beta,\lambda'\lambda)}$ and $G^{(\beta,\lambda'\lambda)}$ are
respectively the hermitean and antihermitean parts of the matrix
$U^{(\beta,\lambda'\lambda)}$ (symmetric and antisymmetric parts,
respectively, in case of a real matrix )
\beqa
F^{(\beta,\lambda'\lambda)}&=&U^{(\beta,\lambda'\lambda)}+U^{(\beta,\lambda'\lambda)\,\dagger}
\,,\\
G^{(\beta,\lambda'\lambda)}&=&U^{(\beta,\lambda'\lambda)}-U^{(\beta,\lambda'\lambda) \,\dagger}
\,.
\eeqa

For the  application of the criterion of Eq.~(\ref{cond}) we
introduce real and imaginary parts of the amplitudes by
\begin{equation}
t_j=x_j+i\,y_j,\quad j=1,...,n\,.
\end{equation}
Since an overall phase is arbitrary, we can take one of $x_j$ or $y_j$ as zero. For
definiteness we set $y_n=0$, so that the amplitude $t_n$ is real. Then introducing the
$2n-1$ dimension real vector $\vec{z}=\{x_1,x_2, ..., x_n,y_1,y_2,...,y_{n-1}\}$, the
observables $f^{(\beta,\lambda'\lambda)}$ and
$g^{(\beta,\lambda'\lambda)}$ can be represented by the following
real quadratic forms
\beqa
f^{(\beta,\lambda'\lambda)}&=&\sum_{ij=1}^{2n-1}z_{i}A^{(\beta,\lambda'\lambda)}_{ij}z_j\,,\\
g^{(\beta,\lambda'\lambda)}&=&\sum_{ij=1}^{2n-1}z_{i}B^{(\beta,\lambda'\lambda)}_{ij}z_j\,,
\eeqa
where the $(2n-1)\times(2n-1)$ matrices
$A^{(\beta, \lambda'\lambda)}$ and $B^{(\beta, \lambda'\lambda)}$
are determined as
\beqa\label{Amatrix}
A^{(\beta,\lambda'\lambda)}&=&\left(
                                   \begin{array}{cc}
                                     F^{(\beta,\lambda'\lambda)} & 0 \\
                                     0 & \widehat{F}^{(\beta,\lambda'\lambda)} \\
                                   \end{array}
                                 \right)\,,\\
B^{(\beta,\lambda'\lambda)}&=&\left(
                                   \begin{array}{cc}
                                     0 & \widehat{G}^{(\beta,\lambda'\lambda)} \\
                                     -\widehat{G}^{(\beta,\lambda'\lambda)T} & 0 \\
                                   \end{array}
                                 \right)\,.\label{Bmatrix}
\eeqa
Here, the matrix $\widehat{F}^{(\beta,\lambda'\lambda)}$ is obtained from
$F^{(\beta,\lambda'\lambda)}$ by canceling the $n$th row and the $n$th column whereas
the matrix $\widehat{G}^{(\beta,\lambda'\lambda)}$ is obtained from
$G^{(\beta,\lambda'\lambda)}$ by canceling the $n$th column. Using the exact expressions
of the matrices $F/G^{(\beta,\lambda'\lambda)}$ one can easily construct the
matrices $A^{(\beta,\lambda'\lambda)}$ and
$B^{(\beta,\lambda'\lambda)}$ of Eqs.~(\ref{Amatrix}) and
(\ref{Bmatrix}), respectively.

Now we study in detail only the simplest case, when the partial wave expansion
of Eq.~(\ref{14}) is truncated at $J_{max}=1/2$. Because of angular momentum
conservation, requiring $|\lambda-\mu|=1/2$ (or $\mu=\lambda/2$), the
total number of amplitudes $t_{\nu\lambda\mu}^{1/2M}$ for each
$\lambda=\pm 1$ is reduced to four. Furthermore, since we exclude from the present consideration
linear photon polarization, only the coefficients with
$\lambda=\lambda'$ appear in the observables in
Eqs.~(\ref{difcrossec}) and (\ref{recoilpol}) according to
Eqs.~(\ref{v0}), (\ref{vc}) and (\ref{walpha}). Then the subsets of
the amplitudes $t_{\nu\lambda\mu}^{1/2M}$ corresponding to $\lambda=1$
and $\lambda=-1$ can be considered separately. This is obvious from
the fact that in this case the observables are combinations either of the
functions $f/g^{(\beta,11)}$ or $f/g^{(\beta,-1-1)}$. Thus we list in
Table~\ref{tab15}  for both  cases the quantum
numbers $M$ and $\nu$ of the amplitudes $t_k\equiv
t_{\nu\lambda\mu}^{1/2M}$ and 
their enumeration  $k=1,\dots, 4$.

\begin{table}[h]
\caption{Enumeration $k$ of the amplitudes
  $t_k\equiv t^{1/2M}_{\nu\lambda\mu}$ for both $\lambda$-values. The
  initial nucleon helicity $\mu$ is 
  fixed by the angular momentum conservation as $\mu=\lambda/2$.}
\begin{ruledtabular}
\begin{tabular}{c|rrrrrrrr}
$k$ & 1 & 2 & 3 & 4 \\
\colrule
$M$ & $-1/2$ & $-1/2$ &  $1/2$ & $1/2$ \\
$\nu$ & $-1/2$ &  $1/2$ & $-1/2$ & $1/2$ \\
\end{tabular}
\end{ruledtabular}
\label{tab15}
\end{table}

 For the case $J_{max}=1/2$ one finds ten values for
$\beta=(I'M';00;jm'm)$ which are listed and 
enumerated by $n_\beta$ from one to ten in Table~\ref{tab16}. One
should note that in 
the absence of target orientation one always has $(IM)=(00)$.
The corresponding linearly independent $4\times 4$ matrices
$F^{(n_\beta,\lambda\lambda)}$ and $G^{(n_\beta,\lambda\lambda)}$ are
listed in Eqs.~(\ref{Vll}) and (\ref{Wll}) of Appendix~\ref{Matrices}.
Those matrices $G^{(n_\beta,\lambda\lambda)}$ which are absent in this listing
are either zero or depend linearly on the matrices in Eqs.~(\ref{Vll}) and
(\ref{Wll}) according to the 
symmetry of Eq.~(\ref{167}). For each $\lambda=\pm 1$ the
set $\{F^{(n_\beta,\lambda\lambda)}, n_\beta=1,\dots,10\}$ forms a
basis in the space of symmetric real $4\times 4$ matrices as does the
set $\{G^{(n_\beta,\lambda\lambda)},n_\beta=2,5,7,8,9,10\}$ in the space of
antisymmetric real $4\times 4$ matrices.
Except for $F^{(1,\lambda\lambda)}$,  all matrices have
a vanishing trace.

\begin{table}[h]
\caption{Enumeration $n_\beta$ of the observables $f/g^{(n_\beta,\lambda\lambda)}$
  for $\beta=(I'M';00;jm'm)$ and $J=1/2$.}
\begin{ruledtabular}
\begin{tabular}{c|cccccccccc}
$n_\beta$ & 1 & 2 & 3 & 4 & 5 & 6 & 7 & 8 & 9 & 10 \\
\colrule
$I'M'$ & 00 &  00 & 00 & 10 &  10 & 10 & 11 & 11 & 11 & 11 \\
$jm'm$ & 000 & 110 & 100 & 000 & 110 & 100 & 000 & 110 & 100 & 1-10 \\
\end{tabular}
\end{ruledtabular}
\label{tab16}
\end{table}

Now, in order to find a complete set of observables (up to already
mentioned possible discrete ambiguities), we have to construct at least one nonsingular
$7\times 7$ matrix using the columns of the matrices
$A^{(n_\beta,\lambda\lambda)}$ of Eq.~(\ref{Amatrix}) and
$B^{(n_\beta,\lambda\lambda)}$ of Eq.~(\ref{Bmatrix}). Because of a
rather simple form of the 
constituent matrices $F^{(n_\beta,\lambda\lambda)}$ and $G^{(n_\beta,\lambda\lambda)}$
(see Appendix~\ref{Matrices}), it is not difficult to
find different combinations of
columns which constitute nonsingular matrices. In fact one can select
almost any set of eight
matrices $A^{(n_\beta,\lambda\lambda)}$ and $B^{(n_\beta,\lambda\lambda)}$.
For example, one may take the columns in the following combination
\begin{equation}
1_{A(1,\lambda)},4_{A(2,\lambda)},3_{A(3,\lambda)},
4_{A(4,\lambda)},3_{B(2,\lambda)},4_{A(5,\lambda)},
1_{B(5,\lambda)},4_{A(6,\lambda)}\,,
\end{equation}
where the notation $k_{A/B(n_\beta,\lambda)}$ means that one selects the $k$th
column from the matrix $A/B^{(n_\beta,\lambda\lambda)}$. Now using for
the differential cross section the
expressions in Eqs.~(\ref{dsig0}) and (\ref{dsigasyc}) for $\sigma^0$
and $\Sigma^c$, respectively, and for the $z$-component of the recoil
polarization $P^0_z$ and $P^c_z$ in Eqs.~(\ref{recoilp0}) and
(\ref{recoilpc}), respectively,
in terms of $f^{(n_\beta,\lambda\lambda)}$ and
$g^{(n_\beta,\lambda\lambda)}$, one finds the following set of 16
observables (see Table~\ref{tab16} for the enumeration
$f/g^{(n_\beta,\lambda\lambda)}$)
\begin{eqnarray}\label{compSet}
\sigma^0,\Sigma^c&:&
f^{(1,\lambda\lambda)},\,f^{(2,\lambda\lambda)},\,f^{(3,\lambda\lambda)},\,g^{(2,\lambda\lambda)},\quad \lambda=\pm 1\,,\nonumber\\
P^0_z,P^c_z&:&
f^{(4,\lambda\lambda)},\,f^{(5,\lambda\lambda)},\,f^{(6,\lambda\lambda)},\,g^{(5,\lambda\lambda)},\quad
\lambda=\pm 1\,,
\end{eqnarray}
where the quantities $f/g^{(n_\beta,\lambda\lambda)}$ are determined by the
Eqs.~(\ref{ReU}) and (\ref{ImU}). Out of this set one may select any 15 observables
for a complete set. As noted, such a set is only a 'minimal' complete
set of observables in the sense, that it
generally determines the required amplitudes up to possible discrete
ambiguities, only. In
other words, if one solves the corresponding system of 15 bilinear equations, one finds
in general more than one solution.

In order to resolve the remaining ambiguities and thus to find a
  proper unique solution, additional information on other observables
  is needed. For a proper selection of additional observables we now 
apply the criterion formulated in
Ref.~\cite{Tabakin}.
Given a linear transformation of the amplitudes $t_k$
\begin{equation}
t_k\to t^{\prime}_k=\sum_{k'k}U_{kk'}t_{k'}\,,
\end{equation}
the criterion of Ref.~\cite{Tabakin} reads as follows:  if there
exists a nontrivial transformation $U$ with the property
\begin{equation}\label{critII}
U^\dag {\cal O} U={\cal O}
\end{equation}
for all matrices ${\cal
O}\in\{F^{(n_\beta,\lambda'\lambda)},G^{(n_\beta,\lambda'\lambda)}\}$
of the set of selected observables, than for any solution  $\{t_k\}$
of this set of observables, the amplitudes 
$\{t'_k\}$ form another solution of the same set, since
\begin{equation}
{\cal O}=\sum_{k'k}t_{k'}^*{\cal O}_{k'k} t_k=\sum_{k'k}t_{k'}^*(U^\dag{\cal
O}U)_{k'k}t_k=\sum_{k'k}t^{\prime
*}_{k'}{\cal O}_{k'k}t^\prime_k\,,
\end{equation}
and thus there is a discrete ambiguity. As is mentioned in
Ref.~\cite{Tabakin} this criterion
is in general not sufficient since it covers only linear transformations $t_k\to t'_k$.
Nevertheless, using this criterion one can resolve at least some of the
possible discrete ambiguities, thus making the general problem easier to solve.

Since our minimal set includes the observable $f^{(1,\lambda\lambda)}$ which is
proportional to the scalar product
\begin{equation}
f^{(1,\lambda\lambda)}=\sum_{k'k}t^*_{k'}F^{(1,\lambda\lambda)}_{k'k}t_k= \frac12
\sum_{k}\left|t_k\right|^2\,,
\end{equation}
the transformations $U$ should preserve the moduli of the
amplitudes. It is therefore natural to consider primarily
unitary $n\times n$ matrices. The
property of Eq.~(\ref{critII}) is then equivalent to the
commutativity of the matrix $U$
with all matrices ${\cal O}$ of the selected set. Application of the criterion in the
present case means, that we have to find a nontrivial transformation $U$ in
the space of unitary $4\times 4$ matrices which commutes with all
matrices $F^{(n_\beta,\lambda'\lambda)}$ and
$G^{(n_\beta,\lambda'\lambda)}$ of the set listed in Eq.~(\ref{compSet}).

Such a matrix $U$ is easily found among the diagonal unitary matrices:
\begin{equation}\label{U1}
U=\left(
           \begin{array}{cccc}
             -1 & \cdot & \cdot & \cdot \\
             \cdot & 1 & \cdot & \cdot \\
             \cdot & \cdot & -1 & \cdot \\
             \cdot & \cdot & \cdot & 1 \\
           \end{array}
         \right)\,.
\end{equation}
At the same time,
it does not commute with any one of the matrices
$F/G^{(n_\beta,\lambda\lambda)}$ for
$n_\beta=7,\dots,10$ (see Table~\ref{tab16}). This means that in order to
resolve the ambiguity under discussion, the minimal set in
Eq.~(\ref{compSet}) should be enlarged by any 
of the observables $f/g^{(n_\beta,\lambda\lambda)}$ belonging to the
recoil polarization components
$P^0_x$ and $P^c_x$ (or $P^0_y$ and $P^c_y$).

It is interesting to note that according to Table~\ref{tab15} the
  matrix of Eq.~(\ref{U1}) corresponds to the transformation of
  Eq.~(\ref{paritytrans1}) 
which in turn is equivalent to the parity exchange of Eq.~(\ref{paritytrans}).
Therefore, the existence of the ambiguity determined by the
transformation $U$ in Eq.~(\ref{U1}) is directly related to
our previous conclusion about the necessity
of measuring the recoil polarization $P_x$ or $P_y$ in order to separate
contributions from states with different parities. In order to resolve
this ambiguity it is sufficient to enlarge the set of observables in
Eq.~(\ref{compSet}), for example,  by $f^{(8,\lambda\lambda)}$. However, there
exists another ambiguity related to another nontrivial diagonal unitary
transformation commuting with this enlarged set, namely
\begin{eqnarray}\label{r23}
\widetilde U=\left(
           \begin{array}{cccc}
             e^{2i\phi_{23}} & \cdot & \cdot & \cdot \\
             \cdot & 1 & \cdot & \cdot \\
             \cdot & \cdot & e^{2i\phi_{23}} & \cdot \\
             \cdot & \cdot & \cdot & 1 \\
           \end{array}
         \right)\,,
\end{eqnarray}
where $\phi_{23}=\phi_2-\phi_3$ is the relative phase of $t_2$ and
$t_3$.

For the elimination of this last
ambiguity one can supplement the existing set by the observable
$g^{(8,\lambda\lambda)}$. It is easy to prove that in the case of
$J_{max}=1/2$ the resulting set of observables turns out to be 'fully'
complete. To show this we firstly note that the magnitudes of all four
amplitudes for $\lambda=1$ is determined by the set of linear
equations, corresponding to the four diagonal matrices
$F^{(n_\beta,11)}$ with $n_\beta=1,3,4,6$ (see Table~\ref{tab16} and
Eq.~(\ref{Vll})). Obviously, the determination of the absolute
squares $|t_i|^2$ from this set does not involve any discrete
ambiguity. Once the magnitudes are known, the relative phases
$\phi_{13}=\phi_3-\phi_1$ and $\phi_{24}=\phi_4-\phi_2$ may be
unambiguously determined using the four matrices
$F/G^{(2,11)}$ and $F/G^{(5,11)}$. This is the only information which may be obtained
from the minimal complete set. For an unambiguous determination of all
four amplitudes we only
need one of the remaining phases $\phi_{23}$ or $\phi_{14}$, since the second one
may always be found from the identity
\begin{equation}\label{phi13}
\phi_{13}-\phi_{23}+\phi_{24}-\phi_{14}=0\,.
\end{equation}
As may be seen from Eq.~(\ref{Vll}), the relative phase $\phi_{23}$ can be extracted from
$F/G^{(8,11)}$.
Obviously, the same procedure can be applied to
the subset $\lambda=-1$. Thus, in order to unambiguously determine the amplitudes
$t^{1/2M_J}_{\nu\lambda\mu}$ the following set of single and double polarization
observables is sufficient
\begin{equation}\label{168}
\sigma^0(I_0)\,,\Sigma^c(I^\odot)\,,P^{0/c}_z(P_{z'},P^\odot_{z'})\,,P_{x_i}^{0/c}(P_{x_i'},P^\odot_{x_i'})\,,
\end{equation}
where for $x_i$ one can take either $x$ or $y$. In Eq.~(\ref{168}) we display in
parentheses the corresponding notation of Ref.~\cite{Roberts} for the observables.

One comment with respect to this result is in order. It is clear that all matrix
elements $t_k$ may be unambiguously determined (apart from an overall phase)
if the modulus of one amplitude, say, for example,
  $|t_1|^2$, and all interference terms $t^*_1t_i$  are known either directly or
  through a chain $t^*_1t_k, t_k^*t_l,\dots, t_m^*t_i$,
  because such interference terms can be expressed as linear combinations of
  observables.  Such a strategy has been discussed and employed in
  Ref.~\cite{ArLeiTomFBS}. In this respect one should note that our
  set, containing the absolute values of all amplitudes is
  overdetermined. The knowledge of $|t_i|^2$ for $i>1$ is not needed
  in this case, since these can be obtained from the obvious identity 
\begin{equation}
|t_i|^2=\frac{t^*_it_j\,t^*_kt_i}{t^*_kt_j}\,,
\end{equation}
for any $j$.
However the structure of our set of equations does not allow the
determination of the magnitude of just only one partial wave amplitude.

\section{Conclusion}\label{conclusion}

The present paper is only the first step towards a systematic approach
to a model
independent partial wave analysis of a complete experiment for the
photoproduction of two pseudoscalar mesons on a nucleon. The scheme,
presented here, is based on a model independent formalism
of a partial wave expansion developed in Ref.\,\cite{FiA12}.

The procedure for finding a complete set is based on two criteria. The
first one, originally developed for deuteron photo- and electrodisintegration in
Refs.~\cite{ArLeiTom,ArLeiTomFBS} allows the elimination of a set of
$2n-1$ independent observables. To partially resolve
possible remaining discrete ambiguities a second criterion from Ref.~\cite{Tabakin} is
used. In the simplest case of truncating the partial wave expansion at
$J_{max}=1/2$ these two criteria turn out to be sufficient
for an unambiguous determination of the eight amplitudes
$t^{1/2M_J}_{\nu\lambda\mu}$. The corresponding complete set
includes beyond the unpolarized cross section,
helicity beam asymmetry, as well as recoil nucleon polarization along the $z$ and one of
the $x$ or $y$ axes with and without circular polarization of the
photon beam. It is rather
interesting, that the complete set of observables for the reactions
discussed here necessarily
includes recoil polarization in the plane orthogonal to the quantization axis. Otherwise,
the contributions of the partial waves with the same total angular momentum $J$ but
different parity cannot be separated. This property distinguishes the
present reaction from those with a single pseudoscalar meson in the
final state, where one can avoid to measure recoil polarization as
demonstrated in Refs.~\cite{Omel,Tiat}.

We are aware of the fact that the practical use of the
present results for $J_{max}=1/2$ is
very limited, since waves with $J=3/2$ appear to be important in both $\pi\pi$ and
$\pi\eta$ channels even in the low energy region. Furthermore, in the
case of truncation at $J_{max}=1/2$
the matrices $F/G^{(n_\beta,\lambda\lambda)}$ are very simple
and an increase of
$J_{max}$ to $3/2$ will probably require not only quantitative but
also some qualitative modifications of the approach. Therefore, the
generalization of this method to higher partial waves will be
considered in a forthcoming paper.

There is,
however, an important conclusion, coming from the present
study. Namely, whereas a complete experiment for a determination
of the spin amplitudes of the reactions considered here is quite 
complicated, because according to the 
analysis of Ref.~\cite{Roberts} it requires the measurement of a
triple polarization observable, the 
truncated partial wave analysis seems to be doable, and thus
further developments in this direction may be very promising.

\section*{Acknowledgment}
This work was supported by the Deutsche Forschungsgemeinschaft (SFB 1044). A.F.
acknowledges additional support by the Dynasty Foundation as well as by RF Federal
programm ``Kadry'' (contract 14.B37.21.0786) and MSE Program 'Nauka' (contract
1.604.2011).

\appendix
\renewcommand{\theequation}{A\arabic{equation}}
\setcounter{equation}{0}
\newpage
\section{Parity separation}\label{parity}

In order to separate the final states of positive parity from those of
negative parity corresponding to the parities of the intermediate
nucleon resonances in the two-step process, we split ${\cal O}^{\lambda
  LJ}_{M}(l_p j_p m_p) $
\beq
{\cal O}^{\lambda LJ}_{M}(l_p j_p m_p) ={\cal O}^{\lambda
  LJ\,+}_{M}(l_p j_p m_p)  +{\cal O}^{\lambda LJ\, -}_{M}(l_p j_p m_p)
\eeq according to the parity $(-1)^{l_p+l_q}$ of the final partial wave $|p \,q;
\big((l_p\frac12)j_p l_q\big)JM\rangle$. Explicitly one finds \beqa {\cal O}^{\lambda
LJ\, \pm}_{M}(l_p j_p m_p)&=& \frac{(-1)^{1+J}\widehat J}{4\sqrt{2\pi}} \sum_{l_q m_q}
((-1)^{l_p+l_q}\pm 1)i^L (-1)^{j_p} \,\widehat l_p\,\widehat j_p\,\widehat l_q\,\widehat
L \, d^{l_q}_{0\,m_q}(\pi/2)\nonumber\\&&\times \left(
\begin{matrix}
j_p& l_q&J \cr m_p&m_q&-M \cr
\end{matrix} \right)
\langle p \,q; \big((l_p\frac12)j_p l_q\big)J||{\cal O}^{\lambda
  L}||\frac{1}{2}\rangle\,.\label{olm-parity}
\eeqa
In the same way we split for $\nu=1/2$ the small $t$-matrix
\beq
t^{JM}_{1/2 \lambda\mu}=t^{JM\,+}_{ \lambda\mu}+t^{JM\,-}_{ \lambda\mu}\,,
\eeq
where the $t^{JM\,\pm}_{ \lambda\mu}$ are defined as in
eq.~(\ref{multipole}) with $\nu=1/2$ and ${\cal O}^{\lambda LJ\, \pm}_{M}$ in place
of ${\cal O}^{\lambda LJ}_{M}$.

For $\nu=-1/2$ one obtains
\beqa
t^{JM}_{-1/2 \lambda\mu}(\omega_1,\omega_2)&=&
(-1)^{\frac{1}{2}}\,\sum_{l_p j_p m_p L}
\left(\begin{array}{ccc} l_p &
    \frac{1}{2} & j_p \cr 0 & \frac{1}{2} & -\frac{1}{2} \cr
\end{array}\right)
\left(\begin{array}{ccc} J & L & \frac{1}{2} \cr \mu-\lambda & \lambda & -\mu\cr
\end{array}\right)
d^{j_p}_{\frac{1}{2}\, m_p}(\pi/2) \,
e^{i(M-m_p)\phi_{qp}}\,\nonumber\\&&\hspace{2cm}
(-1)^{l_p+m_p}\,
{\cal O}^{\lambda LJ}_{M}(l_p j_p m_p)\,,\label{nu-minus}
\eeqa
where we have used the symmetry property of the 3j-symbol and the
property of the small $d$-matrices
\beq
d^j_{m'm}(\pi-\beta)=(-1)^{j-m}d^j_{-m'm}(\beta)
\eeq
yielding for $\beta=\pi/2$
\beq
d^{j_p}_{-\frac{1}{2} m_p}(\pi/2)=(-1)^{j_p-m_p}d^{j_p}_{\frac{1}{2} m_p}(\pi/2)\,.
\eeq
From the same property follows
\beq
(-1)^{m_q}d^{l_q}_{0 m_q}(\pi/2)=(-1)^{l_q}d^{l_q}_{ 0 m_q}(\pi/2)\,,
\eeq
and thus one obtains
\beq
(-1)^{m_p}d^{l_q}_{0 m_q}(\pi/2)
\left(
\begin{matrix}
j_p& l_q&J \cr m_p&m_q&-M \cr
\end{matrix} \right)
=(-1)^{l_q+M}d^{l_q}_{0 m_q}(\pi/2)
\left(
\begin{matrix}
j_p& l_q&J \cr m_p&m_q&-M \cr
\end{matrix} \right)\,.
\eeq
This leads to the relation
\beq
(-1)^{l_p+m_p}{\cal O}^{\lambda LJ}_{M}(l_p j_p m_p) =(-1)^{M}\Big({\cal O}^{\lambda
  LJ\,+}_{M}(l_p j_p m_p)  -{\cal O}^{\lambda LJ\, -}_{M}(l_p j_p m_p) \Big)\,.
\eeq
Inserting this into eq.~(\ref{nu-minus}) leads finally to
\beq
t^{JM}_{-1/2 \lambda\mu}=(-1)^{\frac{1}{2}+M}\Big(t^{JM\,+}_{
  \lambda\mu}-t^{JM\,-}_{ \lambda\mu} \Big)\,.
\eeq
Therefore, the separate parity contributions are given by
\beq
t^{JM\,\pm}_{ \lambda\mu}=\frac{1}{2}\Big(t^{JM}_{1/2
  \lambda\mu}\pm(-1)^{\frac{1}{2}+M} t^{JM}_{-1/2 \lambda\mu} \Big)\,.
\eeq

\renewcommand{\theequation}{B\arabic{equation}}
\setcounter{equation}{0}
\newpage
\section{Listing of the Matrices
  $F/G^{(n_\beta,\lambda\lambda)}$}\label{Matrices}

Here we list the ten symmetric and linearly independent matrices
  $\{F^{(n_\beta,\lambda\lambda)}, n_\beta=1,\dots,10\}$  and the six
  asymmetric matrices
  $\{G^{(n_\beta,\lambda\lambda)},n_\beta=2,5,7,8,9,10\}$  for
  $\lambda=1$. One should consult Table~\ref{tab16} for the correspondence
between $\beta=(I'M';00;jm'm)$ and the enumeration $n_\beta$.
\begin{eqnarray}\label{Vll}
\begin{array}{lll}
F^{(1,11)}=\frac{1}{2}\left(
      \begin{array}{rrrr}
        1 & \cdot & \cdot & \cdot \\
        \cdot & 1 & \cdot & \cdot \\
        \cdot & \cdot & 1 & \cdot \\
        \cdot & \cdot & \cdot & 1 \\
      \end{array}
    \right)\,,&
F^{(2,11)}=-\frac{1}{2\sqrt{2}}\left(
      \begin{array}{rrrr}
        \cdot & \cdot & 1 & \cdot \\
        \cdot & \cdot & \cdot & 1 \\
        1 & \cdot & \cdot & \cdot \\
        \cdot & 1 & \cdot & \cdot \\
      \end{array}
    \right)\,,&
F^{(3,11)}=\frac{1}{2}\left(
      \begin{array}{rrrr}
       -1 & \cdot & \cdot & \cdot \\
        \cdot &-1 & \cdot & \cdot \\
        \cdot & \cdot & 1 & \cdot \\
        \cdot & \cdot & \cdot & 1 \\
      \end{array}
    \right)\,, \\
& &  \\
F^{(4,11)}=\frac{1}{2}\left(
      \begin{array}{rrrr}
       -1 & \cdot & \cdot & \cdot \\
        \cdot & 1 & \cdot & \cdot \\
        \cdot & \cdot &-1 & \cdot \\
        \cdot & \cdot & \cdot & 1 \\
      \end{array}
    \right)\,,&
F^{(5,11)}=\frac{1}{2\sqrt{2}}\left(
      \begin{array}{rrrr}
        \cdot & \cdot & 1 & \cdot \\
        \cdot & \cdot & \cdot & -1 \\
        1 & \cdot & \cdot & \cdot \\
        \cdot & -1 & \cdot & \cdot \\
      \end{array}
    \right)\,,&
F^{(6,11)}=\frac{1}{2}\left(
      \begin{array}{rrrr}
        1 & \cdot & \cdot & \cdot \\
        \cdot & -1 & \cdot & \cdot \\
        \cdot & \cdot & -1 & \cdot \\
        \cdot & \cdot & \cdot & 1 \\
      \end{array}
    \right)\,,\\
& & \\
F^{(7,11)}=-\frac{1}{2\sqrt{2}}\left(
      \begin{array}{rrrr}
        \cdot & 1 & \cdot & \cdot \\
        1 & \cdot & \cdot & \cdot \\
        \cdot & \cdot & \cdot & 1 \\
        \cdot & \cdot & 1 & \cdot \\
      \end{array}
    \right)\,,&
F^{(8,11)}=\frac{1}{2}\left(
      \begin{array}{rrrr}
        \cdot & \cdot & \cdot & \cdot \\
        \cdot & \cdot & 1 & \cdot \\
        \cdot & 1 & \cdot & \cdot \\
        \cdot & \cdot & \cdot & \cdot \\
      \end{array}
    \right)\,,&
F^{(9,11)}=\frac{1}{2\sqrt{2}}\left(
      \begin{array}{rrrr}
        \cdot & 1 & \cdot & \cdot \\
        1 & \cdot & \cdot & \cdot \\
        \cdot & \cdot & \cdot &-1 \\
        \cdot & \cdot &-1 & \cdot \\
      \end{array}
    \right)\,,\\
& & \\
F^{(10,11)}=-\frac{1}{2}\left(
      \begin{array}{rrrr}
        \cdot & \cdot & \cdot & 1 \\
        \cdot & \cdot & \cdot & \cdot \\
        \cdot & \cdot & \cdot & \cdot \\
        1 & \cdot & \cdot & \cdot \\
      \end{array}
    \right)\,\cdot&&\\
\end{array}
\end{eqnarray}
The linearly independent matrices $G^{(n_\gamma,11)}$ are
\begin{eqnarray}\label{Wll}
\begin{array}{lll}
 G^{(2,11)}=\frac{1}{2\sqrt{2}}\left(
     \begin{array}{rrrr}
        \cdot & \cdot & 1 & \cdot \\
        \cdot & \cdot & \cdot & 1 \\
        -1 & \cdot & \cdot & \cdot \\
        \cdot & -1 & \cdot & \cdot \\
      \end{array}
    \right)\,, &
G^{(5,11)}=\frac{1}{2\sqrt{2}}\left(
      \begin{array}{rrrr}
        \cdot & \cdot & -1 & \cdot \\
        \cdot & \cdot & \cdot & 1 \\
        1 & \cdot & \cdot & \cdot \\
        \cdot & -1 & \cdot & \cdot \\
      \end{array}
    \right)\,, &
G^{(7,11)}=\frac{1}{2\sqrt{2}}\left(
      \begin{array}{rrrr}
        \cdot & -1 & \cdot & \cdot \\
        1 & \cdot & \cdot & \cdot \\
        \cdot & \cdot & \cdot & -1 \\
        \cdot & \cdot & 1 & \cdot \\
      \end{array}
    \right)\,,\\
& & \\
G^{(8,11)}=\frac{1}{2}\left(
      \begin{array}{rrrr}
        \cdot & \cdot & \cdot & \cdot \\
        \cdot & \cdot & -1 & \cdot \\
        \cdot & 1 & \cdot & \cdot \\
        \cdot & \cdot & \cdot & \cdot \\
      \end{array}
    \right)\,, &
G^{(9,11)}=\frac{1}{2\sqrt{2}}\left(
      \begin{array}{rrrr}
        \cdot & 1 & \cdot & \cdot \\
        -1 & \cdot & \cdot & \cdot \\
        \cdot & \cdot & \cdot & -1 \\
        \cdot & \cdot & 1 & \cdot \\
      \end{array}
    \right)\,, &
G^{(10,11)}=\frac{1}{2}\left(
      \begin{array}{rrrr}
        \cdot & \cdot & \cdot & -1 \\
        \cdot & \cdot & \cdot & \cdot \\
        \cdot & \cdot & \cdot & \cdot \\
        1 & \cdot & \cdot & \cdot \\
      \end{array}
    \right)\,.\\
\end{array}
\end{eqnarray}
For $\lambda=-1$ and $I=M=0$ the matrices are related to the above
ones by
\beqa
F/G^{(I'M';00;jm'm,-1-1)}&=&(-1)^j\,F/G^{(I'M';00;jm'm,11)}\,.
\eeqa



\begin{thebibliography}{00}

\bibitem{Kashev2pi0}
V.~Kashevarov, {\it et al.}, Phys.\ Rev.\ C {\bf 85}, 064610 (2012).

\bibitem{Roberts}
W.~Roberts and T.~Oed, Phys.\ Rev.\ C {\bf 71}, 055201 (2005).

\bibitem{Workman}
R.~L.~Workman, Phys.\ Rev.\ C {\bf 83}, 035201 (2011).

\bibitem{Tiat}
L.~Tiator, AIP Conf.\ Proc.\  {\bf 1432}, 162 (2012)  [arXiv:1109.0608 [nucl-th]].

\bibitem{Grush}
V.~F. Grushin, in {\it Photoproduction of Pions on Nucleons and Nuclei},
edited by A.~A. Komar, (Nova Science, New York, 1989), p. 1ff.

\bibitem{Omel}
A.~S.~Omelaenko, Sov.\ J.\ Nucl.\ Phys.\ {\bf 34}, 406 (1981).

\bibitem{FiA12}
A.~Fix and H.~Arenh\"ovel, Phys.\ Rev.\ C {\bf 85}, 035502 (2012).

\bibitem{ArLeiTom}
H.~Arenh\"ovel, W.~Leidemann and E.~L.~Tomusiak, Nucl.\ Phys.\ A {\bf
  641}, 517 (1998).

\bibitem{ArLeiTomFBS}
H.~Arenh\"ovel, W.~Leidemann and E.~L.~Tomusiak, Few Body Syst.\  {\bf
  28}, 147 (2000)

\bibitem{Tabakin}
W.-T.~Chiang, and F.~Tabakin, Phys.\ Rev.\ C {\bf 55}, 2054 (1997).

\bibitem{Ros57}
E.~M.~Rose, {\it Elementary Theory of Angular Momentum}, Wiley New York 1957.

\bibitem{ArF13}
H.~Arenh\"ovel  and A.~Fix, to be published. 

\end{thebibliography}
\end{document}